\documentclass[12pt]{article}
\usepackage{amssymb,amsmath,latexsym}
\title{Spectral averaging techniques for Jacobi matrices}

\author{Rafael del Rio$^1$, Carmen Martinez$^1$, Hermann
Schulz-Baldes$^2$
\\
\\
$^1${\small IIMAS, UNAM, Mexico City, Mexico}
\\
$^2${\small
Department Mathematik, Universit\"at Erlangen-N\"urnberg, Germany}
}

\date{ }

\newtheorem{theo}{Theorem}

\newtheorem{proposi}{Proposition}

\newcommand{\CC}{{\mathbb C}}
\newcommand{\NN}{{\mathbb N}}
\newcommand{\RR}{{\mathbb R}}

\newcommand{\PP}{{\bf P}}

\newcommand{\Oo}{{\cal O}}

\newcommand{\Tt}{{\cal T}}

\newcommand{\Hh}{{\cal H}}

\begin{document}

\maketitle


\begin{abstract}
Spectral averaging techniques for one-dimensional discrete Schr\"odinger operators are revisited and extended. In particular, simultaneous averaging over several parameters is discussed. Special focus is put on proving lower bounds on the density of the averaged spectral measures. These Wegner type estimates are used to analyze stability properties for the spectral types of Jacobi matrices under local perturbations.
\end{abstract}

\vspace{.5cm}

\section{Introduction}

Spectral averaging techniques for one-dimensional Sturm-Liouville or Jacobi operators
have been developed and applied in various guises
already for almost four decades (see {\it e.g.}  \cite{Sim0} and references therein).
The basic idea is that the Hamiltonian may depend on some parameters (such as boundary
conditions, coupling constants and alike) and
that the spectral measures
averaged over these parameters are absolutely continuous
w.r.t. the Lebesgue measure.
The main object of this work is to
give various criteria on local perturbations that, on top of that,
guarantee that the Lebesgue measure is absolutely continuous w.r.t. to the averaged spectral
measures, thus showing that they are equivalent.
Applications of this equivalence concern spectral analysis.
In fact, local perturbations may change drastically spectral properties of Jacobi operators,
in particular, properties which are related to the singular part of the spectrum. Nevertheless, if
one knows that some of these properties hold for sets of parameters which have a large measure,
then it is possible to prove results about stability of them.

\vspace{.2cm}

In Section
\ref{sec-formulas} we include the necessary background for the spectral
averaging techniques.
The case of one-parameter spectral averaging discussed in Section \ref{sec-oneparameter}
is a discrete version of results for Sturm-Liouville operators obtained in
\cite{dRT} (see also \cite{dRM}), reformulated using a Birman-Schwinger operator instead of an
associated regular problem.
The results on averaging over several parameter in Section \ref{sec-several}
are related to results of Wegner \cite{Weg}
on the density of states for (also multi-dimensional) random Schr\"odinger operators
which have recently been made rigorous by Hislop and M\"uller \cite{HM}.
However, we do not only deal with homogeneous operators, but allow that the
randomness is only in a finite volume. In terms of the
strength of the disorder we estimate the size of this volume needed in order to
insure that the averaged spectral measure is equivalent to the Lebesgue measure.
We argue heuristically what an optimal estimate on this volume would be
(which we were unable to prove)
and explain how it
would allow to correct the wrong weak-disorder scaling behavior of the bounds
obtained in \cite{Weg,HM}. Let us also cite \cite{BS} for further results on several parameter spectral averaging. Section \ref{sec-applic} exhibits some application to
the spectral analysis of the Jacobi operators studied in
Section \ref{sec-several}.

\section{Recollection of basic formulas}
\label{sec-formulas}

This section is a review of several well-known results, which will
be used below. Let $(t_n)_{n\in\NN}$ and $(v_n)_{n\in \NN}$ be sequences of
respectively positive and real numbers, and $\alpha,\beta\in\RR$.
For a given $N\in\NN$, the finite Jacobi matrix $H_{\alpha,\beta}^N$
with left boundary condition $\alpha\in(-\frac{\pi}{2},\frac{\pi}{2})$ and right boundary
condition $\beta\in(0,\pi)$ is an operator on the finite-dimensional
Hilbert space $\ell^2(\{1,...,N\})$ given by

\begin{equation}
\label{eq-Hfinite}
(H_{\alpha,\beta}\phi)_n
\,=\,t_{n+1}\phi_{n+1}+v_{n}\phi_{n}+t_{n}\phi_{n-1} \, ,
\qquad n=1,\ldots,N\;,
\end{equation}
where $t_1=t_{N+1}=1$, together with the boundary conditions
$$
\sin(\alpha)\phi_1-\cos(\alpha)\phi_0=0\;, \qquad
\sin(\beta)\phi_{N+1}+\cos(\beta)\phi_N=0\;.
$$
The matrix written out explicitely is

\begin{equation}
\label{eq-Hfinite2} H^{N}_{\alpha ,\beta} \;=\; \left(
\begin{array}{ccccccc}
v_1+\tan(\alpha)       & t_2  &        &        &         \\
t_2      & v_2    &  t_3  &        &            \\
              & \ddots & \ddots & \ddots  &        \\
                    &        & t_{N-1} & v_{N-1} & t_N   \\
                    &        &        & t_N  & v_N+\cot(\beta)
\end{array}
\right) .
\end{equation}
Dirichlet boundary conditions are given if $\alpha=0$ and
$\beta=\frac{\pi}{2}$. We will also consider the limit $N\to\infty$
of semi-infinite Jacobi matrices. If $H_{\alpha,\beta}$
is in the Weyl limit point
case at infinity, then there is
a unique self-adjoint limit operator denoted by $H_\alpha$,
independent of $\beta$.

\subsection{Transfer matrices}

The transfer matrices are defined  for any complex energy $z$  as

\begin{equation} \label{eq-transfer} \Tt_n^z \;=\; \left(
\begin{array}{cc}
(z\,-\,v_n)\,t_n^{-1} & - t_n \\
t_n^{-1} & {\bf 0}
\end{array}
\right) \;, \qquad n=1,\ldots,N \;.
\end{equation}
Then we introduce the transfer matrices over several sites by
$$
\Tt^z(n,m)\;=\; \Tt_n^z\cdot\ldots\cdot\Tt^z_{m+1}\;,\qquad n>m \;,
$$
and $\Tt^z(n,n)={\bf 1}$.
They allow to write out all those solutions  of the finite
difference equation $H^N_{\alpha,\beta}\phi^z(\alpha)=z\phi^z(\alpha)$
satisfying the left boundary condition:
\begin{equation}\label{eq-Schr}
\left(
\begin{array}{cc}
t_{n+1}\phi^z_{n+1}(\alpha)\\
\phi_n^z(\alpha)
\end{array}
\right) \;=\; \Tt^z(n,0) \left(
\begin{array}{cc}
\cos(\alpha)\\
\sin(\alpha)
\end{array}
\right) \:.
\end{equation}
The right boundary condition is satisfied precisely at the
eigenvalues of $H^N_{\alpha,\beta}$. Hence $z\in\RR$ is an
eigenvalue of $H^N_{\alpha,\beta}$  if and only if for some
$\lambda\neq 0$
$$
\Tt^z(N,0) \left(
\begin{array}{c}
\cos(\alpha)\\
\sin(\alpha)
\end{array}
\right) \;=\; \lambda\;
\left(
\begin{array}{c}
\cos(\beta)\\
\sin(\beta)
\end{array}
\right) \;.
$$

Let us introduce the following notations for the entries of the
transfer matrix
\begin{equation}\label{eq-transferdef}
\Tt^z(N,0)\;=\; \left(
\begin{array}{cc}
a^z_N & b^z_N \\
c^z_N & d^z_N
\end{array}
\right)\;.
\end{equation}
As first formula, let us recall the result of a Wronskian calculation.

\begin{proposi}
\label{prop-Wronskian} If $\phi_n^z=\phi_n^z(0)$ is the solution
with initial conditions $\phi_1^z=1$ and $\phi_0^z=0$, then
$$
a^z_N\overline{c^z_N}-\overline{a^z_N}c^z_N \;=\;
(z-\overline{z})\,\sum_{n=1}^N|\phi_n^z|^2
 \;.
$$
\end{proposi}

\noindent {\bf Proof.} It follows from \eqref{eq-Schr} and the definition
\eqref{eq-transferdef} that
$$
a^z_N\overline{c^z_N}-\overline{a^z_N}c^z_N \;=\;
t_{N+1}\phi _{N+1}^z\overline{\phi_N^z}-t_{N+1}\overline{\phi _{N+1}^z}\phi_N^z
\,.
$$
Replacing twice the Schr\"odinger equation $t_{N+1}\phi_{N+1}^z=(z-v_N)\phi_N^z-t_N\phi_{N-1}^z$ gives
$$
a^z_N\overline{c^z_N}-\overline{a^z_N}c^z_N \;=\;
(z-\overline{z})|\phi _{N}^z|^2+
t_N\phi _{N}^z\overline{\phi_{N-1}^z}-t_{N}\overline{\phi _{N}^z}\phi_{N-1}^z
\,.
$$
Iteration over $N$ proves the formula.

\hfill $\Box$

\subsection{Finite volume Green's function identities}

The Green's function of $H^N_{\alpha,\beta}$ is defined by
$$
G^N_{\alpha,\beta}(z,n,m) \;=\;\langle
n|(H^N_{\alpha,\beta}-z)^{-1}|m\rangle\;,
$$
where $n,m=1,\ldots,N$ and $\Im m(z)>0$. For Dirichlet boundary
conditions, we drop the indices $\alpha$ and $\beta$. Furthermore,
we set $G^N_{\alpha,\beta}(z)=G^N_{\alpha,\beta}(z,1,1)$. The latter
is linked to the spectral measure $\rho^N_{\alpha,\beta}$ of
$H^N_{\alpha,\beta}$ w.r.t. the state $|1\rangle$ by
$$
G^N_{\alpha,\beta}(z)\;=\;
\int\rho^N_{\alpha,\beta}(dE)\;\frac{1}{E-z}\;.
$$

Some connections of the Green's function to the transfer matrix are
given in the following proposition (other relations can also be
obtained, but will not be used here).

\begin{proposi}
\label{prop-Green} One has
$$
\frac{1}{a^z_N}\;=\; -\,G^N(z,1,N)\;,\qquad \frac{b^z_N}{a^z_N}\;=\;
G^N(z,1,1)\;,\qquad \frac{c^z_N}{a^z_N}\;=\; -\,G^N(z,N,N)\;.
$$
\end{proposi}

\noindent {\bf Proof.} By Cramer's rule
$$
G(z,1,N)\;=\;
(-1)^{N+1}\;\frac{\det(H^N-z)_{1,N}}{\det(H^N-z)}\;,
$$
where $\det(H^N-z)_{1,N}$ is the subdeterminant with the first row and the $N$th column erased.
Evaluation gives  $\det(H^N-z)_{1,N}=t_2\cdots t_N$. Furthermore, developing by the last column, one gets
$$\det(H^N-z)=(v_N-z)\det(H^{N-1}-z)-t_N^2\det(H^{N-2}-z).$$
Hence $N\mapsto \det(H^N-z)$ satisfies the same recurrence relation as
$N\mapsto (-1)^{N}t_2\cdots t_{N+1}\phi^z_{N+1}$
as can be deduced from the Schr\"odinger equation. Moreover, the initial conditions coincide, namely
$\det(H^1-z)=v_1-z=-t_2\phi_2$ and $\det(H^2-z)=(v_1-z)(v_2-z)-t_2^2=(-1)^2t_2t_3\phi_3$. Therefore one deduces that
$\det(H^N-z)=(-1)^{N}t_1\cdots  t_{N+1}\phi^z_{N+1}=(-1)^Nt_1\cdots t_Na_N^z$ and
$$
G(z,1,N)\;=\;
(-1)^{N+1}\;\frac{t_2\cdots t_N}{(-1)^{N}t_2\cdots t_N a_N^z}\;=\;-\;\frac{1}{a_N^z}.
$$
For the second equality, let us start from
$$
G(z,1,1)\;=\;
\frac{\det(\tilde{H}^N-z)}{\det(H^N-z)}\;,
$$
where $\tilde{H}^N$ is the $(N-1)\times(N-1)$ matrix obtained
from $H^N$ by restriction to the sites $2,\dots ,N$. From the above calculation, we have that
$(-1)^{N-1}t_2\cdots t_N\tilde{a}_N^z=\det(\tilde{H}^N-z)$ where $\tilde{a}_N^z$ is the upper left entry of the transfer
matrix $\Tt_N^z\cdots \Tt^z_{2}$. By multiplication with $\Tt_1^z$ one readily verifies that $\tilde{a}_N^z=-b_N^z$. Thus
replacing $\det(H^N-z)=(-1)^Nt_1\cdots t_Na_N^z$ gives
$$
G(z,1,1)\;=\;
\frac{b_N^z}{a_N^z}
\;.
$$
Finally, again by Cramer's rule,
$$
G(z,N,N)\;=\;
\frac{\det(H^{N-1}-z)}{\det(H^N-z)}
\;.
$$
Using again twice the identity  $\det(H^N-z)=(-1)^{N}t_1\cdots t_{N+1}\phi^z_{N+1}$ allows to conclude
the proof of the last identity.
\hfill $\Box$

\begin{proposi}
\label{prop-Greenboundarycond} The dependence of the Green's
function $G^N_{\alpha,\beta}(z)$ on the boundary conditions is given
by
$$
G^N_{\alpha,\beta}(z)\;=\;\frac{b^z_N-d^z_N\cot(\beta)}{
a^z_N+b^z_N\tan(\alpha)-c^z_N\cot(\beta)-d^z_N\tan(\alpha)\cot(\beta)}\;.
$$
\end{proposi}

\noindent {\bf Proof.} The boundary conditions $\alpha,\beta$ can be incorporated in the potential
values $v_1,v_N$ as in \eqref{eq-Hfinite2}, which is then understood to have Dirichlet boundary conditions.
The resulting transfer matrix from $1$ to $N$ can be expressed in terms of the transfer matrix $\Tt^z(N,0)$:
\begin{equation}
\label{eq-boundaryincorp}
\left(
\begin{array}{cc}
1 & -\cot(\beta) \\
0 & 1
\end{array}
\right)
\left(
\begin{array}{cc}
a^z_N & b^z_N \\
c^z_N & d^z_N
\end{array}
\right)
\left(
\begin{array}{cc}
1 & 0 \\
\tan(\alpha) & 1
\end{array}
\right)
\;.
\end{equation}
Evaluating and extracting the upper left and right entries concludes the proof
together with the second formula of Proposition \ref{prop-Green}.
\hfill $\Box$

\vspace{.2cm}

\begin{proposi}
\label{prop-Greenaveraged} For $\Im m(z)>0$,
$$
\int^\pi_0\frac{d\beta}{\pi}\; G^N_{\alpha,\beta}(z)\;=\;
\frac{b^z_N+\imath d^z_N}{(a^z_N+b^z_N\tan(\alpha))+\imath(c^z_N+d^z_N\tan(\alpha))}\;,
$$
and for $E\in\RR$
\begin{equation}
\label{eq-Greenav}
\lim_{\epsilon\to 0}\;\Im m\;
\int^\pi_0\frac{d\beta}{\pi}\; G^N_{\alpha,\beta}(E+\imath \epsilon)\;=\;
\frac{1}{|a^E_N+b^E_N\tan(\alpha)|^2+|c^E_N+d^E_N\tan(\alpha)|^2}\;.
\end{equation}
\end{proposi}

\noindent {\bf Proof.} Using \eqref{eq-boundaryincorp} it is easy to deduce the
formulas for arbitrary boundary $\alpha$ from the case $\alpha=0$. Hence it is
sufficient to consider the latter case. From Proposition
\ref{prop-Greenboundarycond} and a change of variables it follows that
$$
\int^\pi_0\frac{d\beta}{\pi}\; G^N_{0,\beta}(z)\;=\;
\int^\pi_0\frac{d\beta}{\pi}\;
\frac{b^z_N\tan(\beta)-d^z_N}{a^z_N\tan(\beta) -c^z_N}\;=\;
\int^\infty_{-\infty}\frac{dx}{\pi(1+x^2)}\;
\frac{b^z_Nx-d^z_N}{a^z_Nx -c^z_N}
\;.
$$
The latter integral can be evaluated by a contour integral. The poles
of the integrand are at $x=\imath,-\imath,\frac{c^z_N}{a^z_N}$. As
$\frac{c^z_N}{a^z_N}=-G^N(z,N,N)$ is in the lower half-plane, the only
pole in the upper half-plane is $x=\imath$. Hence the residue theorem
directly implies the first formula of the proposition. The second one
follows directly by calculating the imaginary part and using the fact
that the coefficients $a^E_N,b^E_N,c^E_N,d^E_N$ are real and satisfy
$a^E_Nd^E_N-b^E_Nc^E_N=1$.

\vspace{.2cm}

It follows immediately from \eqref{eq-Greenav} and the de la Vall\'ee-Poussin
theorem that $\int^\pi_0
\frac{d\beta}{\pi} \,\rho^N_{\alpha,\beta}$ is absolutely continuous with
density given by the r.h.s. of \eqref{eq-Greenav}.
\hfill $\Box$

\subsection{Pr\"ufer variables}

In this section, the energy is real and hence we set $z=E\in\RR$.
For any fixed left boundary condition $\alpha$, we define as
\cite{JSS} the Pr\"ufer phases $\theta^{E}_{n}(\alpha)$ and Pr\"ufer
radius $R^{E}_{n}(\alpha)$ by

\begin{equation}
\label{eq-Pruefer}
R^{E}_n\left(
\begin{array}{c}
\cos(\theta^{E}_{n})\\
\sin(\theta^{E}_{n})
\end{array}
\right) \;=\; \Tt^E(n,0)\; \left(
\begin{array}{c}
\cos(\alpha) \\
\sin(\alpha)
\end{array}
\right)\;,
\end{equation}
together with the condition $-\frac{\pi}{2}<\theta^E_{n+1}(\alpha)
-\theta^E_n(\alpha)<\frac{3\pi}{2}$ and $\theta^E_0(\alpha)=\alpha$.
Next we derive a few formulas used in the sequel.

\begin{proposi}
\label{prop-alphaaverage}
$$
\int^{\pi}_{0}\frac{d\alpha}{\pi}\;\frac{1}{R^{E}_N(\alpha)^2}
\;=\;1\;.
$$
\end{proposi}

\noindent {\bf Proof.}
Setting $\Tt=\Tt^E(N,0)$ and $e_\alpha=\left(
\begin{array}{c}
\cos(\alpha) \\
\sin(\alpha)
\end{array}
\right)$,
the integral is given by
$$
\int^{\pi}_{0}\frac{d\alpha}{\pi}\;\frac{1}{R^{E}_N(\alpha)^2}
\;=\;
\int^{\pi}_{0}\frac{d\alpha}{\pi}\;\frac{1}{
\langle e_\alpha|\Tt^*\Tt|e_\alpha\rangle}
\;.
$$
Now $\Tt^*\Tt$ is a positive matrix of determinant $1$, hence its
eigenvalues are $\kappa,\frac 1 \kappa >0$ and it is diagonalized
by an orthogonal matrix. As $d\alpha$ is rotation invariant, it follows
that
$$
\int^{\pi}_{0}\frac{d\alpha}{\pi}\;\frac{1}{R^{E}_N(\alpha)^2}
\;=\;
\int^{\pi}_{0}\frac{d\alpha}{\pi}\;\frac{1}{
\kappa \cos^2(\alpha)+\frac 1 \kappa \sin^2(\alpha)}
\;=\;
\int^{\infty}_{-\infty}\frac{dx}{\pi}\;\frac{1}{
\kappa +\frac 1 \kappa x^2}
\;=\;1
\;,
$$
which concludes the proof.
\hfill $\Box$

\vspace{.2cm}

The first of the following two formulas was already proven in \cite{JSS}.

\begin{proposi}
\label{prop-phasederivatives} {\rm (i)} One has
$$
R^{E}_N(\alpha)^2\;\partial_E \theta^{E}_N(\alpha) \;=\;
-\,\sum_{n=1}^N\,|\phi^{E}_n(\alpha)|^2\;.
$$
{\rm (ii)} For the derivative w.r.t. the potential value $v_n$
with $n\leq N$, one
has
$$
R^{E}_N(\alpha)^2\;\partial_{v_n} \theta^{E}_N(\alpha) \;=\;
|\phi^{E}_n(\alpha)|^2\;.
$$
\end{proposi}

\noindent {\bf Proof.} Due to the special form \eqref{eq-transfer} of the
transfer matrices, item (i) follows directly from item (ii). Hence we focus
on (ii). Furthermore, let us suppress the $\alpha$ in all notations.
Deriving $\tan(\theta^E_N)=\frac{\phi^E_N}{t_{N+1}\phi^E_{N+1}}$ w.r.t. to $v_n$
gives
$$
\partial_{v_n} \theta^{E}_N(\alpha)
\;=\;
\frac{1}{1+\bigl(\frac{\phi^E_N}{t_{N+1}\phi^E_{N+1}}\bigr)^2}
\;
\partial_{v_n} \frac{\phi^E_N}{t_{N+1}\phi^E_{N+1}}
\;.
$$
Evaluation and using $(R^{E}_N)^2=
(\phi^E_N)^2+(t_{N+1}\phi^E_{N+1})^2$ gives
$$
\partial_{v_n} \theta^{E}_N(\alpha)
\;=\;
\frac{1}{(R^{E}_N)^2}
\left[
(\partial_{v_n} \phi^E_N)(t_{N+1}\phi^E_{N+1})
-
(\phi^E_N)(\partial_{v_n} t_{N+1}\phi^E_{N+1})
\right]
\;.
$$
Now, as long as $n<N$, the term in the brackets can be evaluated by replacing twice the
Schr\"odinger equation
$t_{N+1}\phi^E_{N+1}=(E-v_n)\phi^E_{N}-t_{N}\phi^E_{N-1}$. At the first step one obtains
$$
\partial_{v_n} \theta^{E}_N(\alpha)
\;=\;
\frac{1}{(R^{E}_N)^2}
\left[
(\partial_{v_n} \phi^E_{N-1})(t_{N}\phi^E_{N})
-
(\phi^E_{N-1})(\partial_{v_n} t_{N}\phi^E_{N})
\right]
\;,
$$
and iteration gives
$$
\partial_{v_n} \theta^{E}_N(\alpha)
\;=\;
\frac{1}{(R^{E}_N)^2}
\left[
(\partial_{v_n} \phi^E_{n})(t_{n+1}\phi^E_{n+1})
-
(\phi^E_{n})(\partial_{v_n} t_{n+1}\phi^E_{n+1})
\right]
\;.
$$
As $\partial_{v_n} \phi^E_{n}=0$ and
$\partial_{v_n} t_{n+1}\phi^E_{n+1}= -\phi^E_{n}$ by the Schr\"odinger equation, one
can conclude the proof.
\hfill $\Box$

\vspace{.2cm}

Finally let us prove Carmona's formula \cite{CL} (which was rediscovered by Pearson
\cite{Pea} and proven by Simon in the discrete case \cite{Sim}).

\begin{proposi}
\label{prop-Carmona} For any $E_0<E_1$,
$$
\frac{1}{2}\left[\rho_{\alpha}([E_0,E_1])+ \rho_{\alpha}((E_0,E_1))
\right]
\;=\;
\lim_{N\to\infty}\;\int^{E_1}_{E_0}\frac{dE}{\pi}\;\frac{\cos^2(\alpha)}{R^{E}_N(\alpha)^2}
\;.
$$
\end{proposi}

\noindent {\bf Proof.} Integrating \eqref{eq-Greenav} w.r.t. energy, one gets
$$
\frac{\pi}{2}
\int^\pi_0\frac{d\beta}{\pi}\,
\left[\rho_{\alpha,\beta}^N([E_0,E_1])+ \rho_{\alpha,\beta}^N((E_0,E_1))
\right]
\;=\;
\int^{E_1}_{E_0}dE\;\frac{\cos^2(\alpha)}{R^{E}_N(\alpha)^2}
\;,
$$
where $\rho_{\alpha,\beta}^N$ is the spectral measure of $H_{\alpha,\beta}^N$
w.r.t. $|1\rangle$. Now let us take the limit $N\to\infty$ of this equation.
On the l.h.s. one may invoke the dominated convergence theorem in order to take
the limit under the integral. As $\rho_{\alpha,\beta}^N$ converges weakly to
$\rho_\alpha$ for all $\beta$, the result follows immediately.
\hfill $\Box$

\section{One-parameter spectral averaging}
\label{sec-oneparameter}

Let us first recall the well-known result on averages
over the left boundary condition \cite{CL,Sim0}. One considers
$\alpha\in[0,\pi) \mapsto H_\alpha$ with corresponding spectral
measures $\rho_\alpha$. Then the spectral averaged measure
$\rho=\int^\pi_0 d\alpha\,\frac{1}{\cos^2(\alpha)}\,\rho_\alpha$
is equal to the Lebesgue measure (up to a factor $\pi$). In fact,
Propositions~\ref{prop-alphaaverage} and
\ref{prop-Carmona} immediately allow to deduce this statement.
As shows the form \eqref{eq-Hfinite2} of the Hamiltonian, this
result can be understood as an average over the potential
value at site $1$  w.r.t. a particular density.
As shown in the last remark of this section, averages
over another single potential value (say at site $N$) can also be
analyzed. We shall be
interested in studying spectral averages in situations where
several entries in the Jacobi matrix are varied. In this section,
this is done with only one parameter, then in the next one with
several parameters.

\vspace{.2cm}

In order to single out the main mechanism,
let us start with a more abstract statement on
spectral averaging. Let $I=[\mu_0,\mu_1]$
be a finite interval of parameters and suppose given a
differentiable family $\mu\in I\mapsto H(\mu)$ of semi-infinite
Jacobi matrices with Dirichlet boundary conditions $\alpha=0$ and
such that only the first $N$ potential values
$v_1,\ldots,v_N$ and off-diagonal terms $t_2,\ldots,t_N$ depend on
$\mu$. The associated spectral measure is denoted by $\rho_\mu$, and
the Pr\"ufer phases by $\theta^{E}_N(\mu)$. Furthermore define the
averaged spectral measure $\rho$ by
\begin{equation}
\label{eq-avmeas}
 \rho\;=\;\int^{\mu_1}_{\mu_0} d\mu\;\rho_\mu \;.
\end{equation}

\begin{proposi}
\label{prop-oneparameter} Suppose that for all $E\in [E_0,E_1]$:

\vspace{.1cm}

\noindent {\rm (i)} $\mu\in I\mapsto \theta^{E}_N(\mu)$ is strictly
monotonous with bounded derivative,

\vspace{.1cm}

\noindent {\rm (ii)} $|\theta^{E}_N(\mu_1)-\theta^{E}_N(\mu_0)|>\pi$.

\vspace{.2cm}

\noindent Then, in the interval $[E_0,E_1]$, the averaged spectral
measure $\rho$ is equivalent to the Lebesgue measure.
\end{proposi}

\noindent {\bf Proof.} Using Proposition~\ref{prop-Carmona}, the
dominated convergence theorem and Fubini's theorem, one gets
$$
\rho([E_0,E_1])\;=\;\lim_{L\to\infty}\;\int^{E_1}_{E_0}dE
\int_{\mu_0}^{\mu_1}d\mu \;\frac{1}{(R_L^{E}(\mu))^2}\;.
$$
Next let us set for $L>N$:
$$
{R}^{E}_{L,N}(\theta)\;=\; \left\| \Tt^{E}(L,N)
\left(\begin{array}{c}
\cos(\theta) \\
\sin(\theta)
\end{array}
\right) \right\|\;.
$$
This does not depend on $\mu$, and
\begin{equation}
\label{eq-splitting}
R^{E}_L(\mu)\;=\;{R}^{E}_{L,N}(\theta^{E}_N(\mu))\;R^{E}_N(\mu) \;.
\end{equation}
Now the continuity of $\mu\mapsto H(\mu)$ gives the bounds
$$
0\;<\;C_0\;\leq\;R^{E}_N(\mu)^2 \;\leq \;C_1\;<\;\infty\;,
$$
which in turn imply
$$
\rho([E_0,E_1])\;\approx\;\lim_{L\to\infty}\;\int^{E_1}_{E_0}dE
\int_{\mu_0}^{\mu_1}d\mu
\;\frac{1}{({R}_{L,N}^{E}(\theta^{E}_N(\mu)))^2}\;,
$$
where the notation $f\approx g$ means that there are positive constants $c_0,c_1$ such that $c_0f\leq g\leq c_1f$, and the limit may not exist and is either
the superior or inferior limit depending on whether one deals with the
upper or lower bound. 
By hypothesis (i), one can make a change of variables in the $\mu$-integral:
$$
\rho([E_0,E_1])\;\approx\;\lim_{L\to\infty}\;\int^{E_1}_{E_0}dE
\int_{\theta(\mu_0)}^{\theta(\mu_1)}d\theta\;
\left|\frac{d\mu}{d\theta}\right|
\;\frac{1}{({R}_{L,N}^{E}(\theta))^2}\;.
$$
Now hypothesis (i) allows to bound the Jacobian from above and below, and then
hypothesis (ii) allows to complete the proof using
Proposition~\ref{prop-alphaaverage}.
\hfill $\Box$

\vspace{.2cm}

This proposition cannot be used to rederive the classical spectral averaging over boundary conditions because the family $\alpha\mapsto H_\alpha$ is not differentiable at $\frac{\pi}{2}$. In our application, we consider Jacobi matrices of the form
$$
H(\mu)\;=\;H(0)+\mu \,W\;,
\qquad
W\;=\;\sum_{n=1}^N w_n\,|n\rangle\langle n|\;,
$$
where $H(0)$ is a given semi-infinite Jacobi matrix, $N<\infty$ and
$w_1,\ldots,w_N\geq 0$ so that the potential $W$ is positive.
Averaging will be done over the parameter $\mu$ and the
following results tell
us under which conditions on the size of the interval
$[\mu_0,\mu_1]$ the above proposition can be applied. One condition
will be expressed in terms of the associated Birman-Schwinger operator
(a self-adjoint $N\times N$ matrix) defined for any energy $E\in\RR$
not in the spectrum of $H^N(0)$ by
$$
K^N_E\;=\; W^{\frac{1}{2}}\,(E-H^N(0))^{-1}\,W^{\frac{1}{2}} \;,
$$
where $H^N(\mu)$ is the $N\times N$ matrix given by the upper left corner of $H(\mu)$.
Let $E_1(\mu)<\ldots<E_N(\mu)$ denote the eigenvalues of $H^N(\mu)$ which are known to be all distinct.

\begin{theo}
\label{theo-secondexample} Fix some $E\in \RR$.
Suppose that two consecutive values $w_m,w_{m+1}$ are strictly positive and that
one of the following conditions on the size of the interval $[\mu_0,\mu_1]$ hold:

\vspace{.1cm}

\noindent {\rm (a)} There are $\mu_0',\mu_1'\in (\mu_0,\mu_1)$ such that
$E_n(\mu_0')=E_{n-1}(\mu_1')=E$ for some $n=2,\ldots,N$.

\vspace{.1cm}

\noindent {\rm (b)}
The potential $W$ is strictly positive, $E$ is not in the spectrum of $H^N(0)$ and
there exist two non-vanishing eigenvalues $\lambda_0(E)$ and $\lambda_1(E)$ of
$K^N_E$ such that
\begin{equation}
\label{eq-cond}
\mu_0\;<\;\frac{1}{\lambda_0(E)}
\;<\;\frac{1}{\lambda_{1}(E)}\;<\;\mu_1 \;.
\end{equation}
Then the averaged spectral measure $\rho$ defined as in {\rm
\eqref{eq-avmeas}} is equivalent to the Lebesgue measure in an open
interval around $E$.
\end{theo}

\noindent {\bf Proof.} The monotonicity condition (i) of
Proposition~\ref{prop-oneparameter} holds due to
Proposition~\ref{prop-phasederivatives}~(ii) applied to all sites,
in particular the two sites $m,m+1$ give the strict monotonicity
because the wave function $\phi^E$ cannot vanish at two consecutive
sites. Now each of the two hypothesis (a) and (b) imply the
condition Proposition~\ref{prop-oneparameter}(ii). In the case (a) this
follows immediately from the
Sturm oscillation theorem ({\it e.g.} Section 3.2 of \cite{JSS}), which
states that the
Pr\"ufer phase $\theta^{E}_N(\mu)$ has to vary by more than $\pi$ in
$[\mu_0,\mu_1]$, namely $\theta^{E}_N(\mu_1)-\theta^{E}_N(\mu_0)>\pi$.
In order to use (b),
let us first rewrite the eigenvalue equation $H^N(\mu)\phi=E\phi$ as
$(E-H^N(0))\phi=\mu W \phi$. As $W$ and therefore $W^{\frac{1}{2}}$ are
invertible, it hence follows that $E$ is an eigenvalue of
$H^N(\mu)$ if and only if $\frac{1}{\mu}$ is an eigenvalue of
$K^N_E$ with eigenstate $W^{\frac{1}{2}}\phi$. This will be used
in order to analyze the eigenvalues of $H^N(\mu)$. If
$I=[\mu_0,\mu_1]$ satisfies \eqref{eq-cond}, then the interval $I$
contains $\mu_0'<\mu_1'$ such that $E_n(\mu_0')=E$ and
$E_{n-1}(\mu_1')=E$ for some $n$ in $2,\ldots N$ and one can apply
the condition (a).
By continuity, all the above holds for an open interval containing
$E$, so that Proposition~\ref{prop-oneparameter} can be applied in order
to conclude the proof.
\hfill $\Box$

\vspace{.2cm}

In order to show that condition (b) invoking the Birman-Schwinger operator
can be more practical, let us treat an

\vspace{.2cm}

\noindent {\bf Example:} Let $N=2$, $w_1,w_2>0$ and $H^2(0)=
\left( \begin{array}{cc} 0 & 1 \\ 1 & 0
\end{array} \right)$. For $E\neq \pm 1$, one then has
$K^2_E=\frac{1}{E^2-1}\left( \begin{array}{cc} Ew_1 & \sqrt{w_1w_2} \\
\sqrt{w_1w_2} & Ew_2 \end{array} \right)$. The eigenvalues are
$$
\lambda_\pm(E)\;=\;
\;\frac{1}{4(E^2-1)}\,
\left[
\;E(w_1+w_2)\pm\sqrt{E^2(w_1^2+w_2^2)+4w_1w_2}
\;\right]
\;.
$$
The condition \eqref{eq-cond} can be written out explicitly,
the order of $\lambda_+(E)$ and $\lambda_-(E)$ therein
depending on whether $|E|>1$ or $|E|<1$.

\vspace{.2cm}

\noindent {\bf Remark} The condition that two adjacent values $w_m,w_{m+1}$
are strictly positive is needed in order to assure the monotonicity condition
(i) in Proposition~\ref{prop-oneparameter}. This can be relaxed by asking,
for example, that $\phi_m^E\neq 0$ for the smallest $m$ such that $w_m>0$.
Note that $\phi_m^E$ then is independent of $W$.

\vspace{.2cm}

\noindent {\bf Remark} The situation where $W$ has only one non-vanishing
entry, say $w_N=1$, can be dealt with in a manner similar to the average
over the boundary condition, that is, one needs to average over
$(\mu_0,\mu_1)=\RR$ and then $\rho=\int_\RR d\mu\,\rho_\mu$ dominates
the Lebesgue measure on all $\RR$. In order to show this, let us decompose
the Pr\"ufer radius as in the proof of Proposition~\ref{prop-oneparameter}
in $R_L^E=R_{L,N}^E(\theta^E_{N})R_{N}^E$. Now $\cot(\theta_N^E)= E -\mu -(t_N^E)^2 \tan(\theta^E_{N-1})$ and $\theta^E_{N-1}$ is independent of $\mu$. Therefore, as
$\mu$ varies in all $\RR$, $\theta^E_{N}$ varies over all $[0,\pi)$ for all
$\theta^E_{N-1}$. Moreover, diagonalizing $R_\eta^* |\Tt^E(L,N)|^2R_\eta
=$diag$(\kappa,1/\kappa)$ with the adequate rotation $R_\eta$ then shows
$$
\int_{\RR}d\mu
\;\frac{1}{({R}_{L,N}^{E}(\theta^{E}_N(\mu)))^2}
\;=\; \int_{\RR}d\mu
\;\frac{1}{\kappa \cos^2(\theta^E_N+\eta)+\frac 1 \kappa
\sin^2(\theta^E_N+\eta)}\;
$$
The integral on the r.h.s. can be bounded below by a constant uniformly
in $\theta^E_{N-1}$ because $\frac{d\mu}{d\theta}\geq 1$. This implies the claim as in the proof of Proposition~\ref{prop-oneparameter}.

\section{Several-parameter spectral averaging}
\label{sec-several}

There are models for which several parameters are available for
spectral averaging, but not any of them is sufficient by itself in
order to lead to an averaged spectral measure which is equivalent to
the Lebesgue measure. Then an averaging over several parameters may
nevertheless allow to prove such a statement. In this section, we
consider the concrete example of a semi-infinite Jacobi matrix of the form
\begin{equation}
\label{eq-mod2}
H_{\lambda,N,v}\;=\;\Delta_N\,+\,\lambda\,\sum_{n=1}^N\;
v_n\;|n\rangle\langle n|\,+\,J_N
\;,
\end{equation}
where $\Delta_N$ is the discrete Laplacian up to site $N$ (namely, $t_n=1$ and
$v_n=0$ for $n=1,\ldots,N$, and vanishing coefficients afterwards), $J_N$ is
an arbitrary Jacobi matrix in the limit point case
with $t_n=v_n=0$ for $n=1,\ldots,N$,
$\lambda\geq 0$ is a coupling constant, and the entries of
$v=(v_1,\ldots,v_N)$ are independent real random variables, each
distributed according to the Lebesgue measure on
$[-\frac{1}{2},\frac{1}{2}]$. We also write $\PP(dv)$ for this
product measure on the unit cube
$I_N=[-\frac{1}{2},\frac{1}{2}]^{\times N}$.
By our methods and a little more notational effort (which we choose
to avoid), the model
\eqref{eq-mod2} could be generalized in order to allow for an arbitrary
periodic background instead of $\Delta_N$ and arbitrary local
perturbations on each periodicity interval.
We use Dirichlet boundary conditions $\alpha=0$ and
suppress the argument $\alpha$ in all formulas below.
The spectral measure of
$H_{\lambda,N,v}$ w.r.t. $|1\rangle$ will be denoted by
$\rho_{\lambda,N,v}$.

\begin{theo}
\label{theo-Wegner} Let $0<\lambda<4$ and $I$ be an open interval
such that its closure is contained in
$(-2+\frac{\lambda}{2},2-\frac{\lambda}{2})$.
Then there exists
an $N=N(\lambda)$ such that the averaged spectral measure
$$
\rho_{\lambda,N}\;=\;\int_{I_N}\PP(dv)\;\rho_{\lambda,N,v} \;,
$$
is equivalent to the Lebesgue measure in $I$.
\end{theo}

This result is very similar to Wegner's upper and lower bound on the
density of states for multi-dimensional discrete random
Schr\"odinger operators \cite{Weg,HM}, however, our proof uses a
different change of variables and is
restricted to the one-dimensional situation
due to the use of Pr\"ufer variables. Our initial intent was to
improve on Wegner's lower bound in the one-dimensional situation in
the weak coupling limit. Indeed, the lower bound on the density of
states as obtained in \cite{Weg,HM} as well as ours goes to zero as
$\lambda$ goes to zero, which is absurd for energies in the spectrum
(the integrated density of states varies only of order $\lambda$ for
all energies in the spectrum \cite{JSS}). The problem behind this
short-coming is that not sufficiently many
random potential values are used for
averaging (because the errors cannot be controled).
As we will argue heuristically below, the
$N$ needed in Theorem~\ref{theo-Wegner} should actually be of order
$\lambda^{-2}$ (which is the localization length).

\vspace{.2cm}

The Pr\"ufer phases and radii at disorder configuration $v$ and
coupling constant $\lambda>0$ are denoted by
$\theta^E_{\lambda,n}(v)$ and $R^E_{\lambda,n}(v)$. The proof of
Theorem~\ref{theo-Wegner} will use modified Pr\"ufer variables (even
though not in an optimized way as explained below). Let us recall their definition,
{\it e.g.} from \cite{PF,JSS}. For $E\in(-2,2)$ and $k=\arccos(E/2)\in
(0,\frac{\pi}{2})$,
one sets
$$
M^E\;=\; \frac{1}{\sqrt{\sin(k)}}\;\left( \begin{array}{cc} \sin(k) & 0 \\
-\cos(k) & 1
\end{array} \right)\;.
$$
Furthermore denote $e_\theta=\left(
\begin{array}{c} \cos (\theta)
\\ \sin(\theta) \end{array} \right)$. Now one defines a
smooth function $m^E:\RR\to\RR$ with
$m^E(\theta+\pi)=m^E(\theta)+\pi$ and $0 < C_1 \le  (m^E)' \le C_2 <
\infty$, by

$$
r(\theta)e_{m^E(\theta)}=M^Ee_{\theta}, \qquad r(\theta)>0 \mbox{ ,
} \qquad m^E(0)\in[-\pi,\pi) \mbox{ . }
$$

\noindent Then the $E$-modified Pr{\"u}fer variables
$(\hat{R}^{E}_{\lambda,n}(v),\hat{\theta}^{E}_{\lambda,n}(v))\in
\RR_+\times\RR$ for initial condition
$\hat{\theta}^E_0=m^E(\theta^E_0)$ are given by
\begin{equation}
\label{eq-prufer1} \hat{\theta}^E_{\lambda,n}(v)\;=\;
m^E(\theta^{E}_{\lambda,n}(v)) \mbox{ , }
\end{equation}
\noindent and
\begin{equation}
\label{eq-prufer2} \left(\begin{array}{c} \hat{R}^{E}_{\lambda,n}(v)
\cos (\hat{\theta}^{E}_{\lambda,n}(v))
\\
\hat{R}^{E}_{\lambda,n}(v) \sin(\hat{\theta}^{E}_{\lambda,n}(v))
\end{array}
\right) \;=\; M^E \left( \begin{array}{c} t_n\, \phi^E_n \\
\phi^E_{n-1}
\end{array} \right)
\mbox{ . }
\end{equation}

\noindent Important for our purposes are two facts ({\it e.g.} \cite{JSS}).
First of all,
$|\hat{\theta}^{E}_{\lambda,n}(v)-\theta^{E}_{\lambda,n}(v)|\leq
2\pi$ for all $n\geq 0$. Second of all, the
behavior of the $E$-modified Pr\"ufer variables is very simple at
$\lambda=0$, namely
\begin{equation}
\label{eq-modPr}
\hat{R}^{E}_{0,n}(v)=\hat{R}^{E}_{0,0}(v)\;,\qquad
\hat{\theta}^{E}_{0,n}(v)= \hat{\theta}^E_0+nk\;.
\end{equation}

\vspace{.2cm}

\noindent {\bf Proof of Theorem~\ref{theo-Wegner}.} The upper bound
follows immediately as in Proposition~\ref{prop-oneparameter}, so we
will only focus on the lower bound here. Let $E_0,E_1\in I$.
Proceeding exactly as in
the proof of Proposition~\ref{prop-oneparameter}, one shows
$$
\rho_{\lambda,N}([E_0,E_1])\;=\; \lim_{L\to\infty}\;\int^{E_1}_{E_0}dE
\int_{I_N}\PP(dv) \; \frac{1}{R_{\lambda,L}^E(v)^2}\; .
$$
The main advantage of this formula is that one can pass to
$E$-modified Pr\"ufer variables at every energy $E\in[E_0,E_1]$.
Indeed, the Pr\"ufer radii are changed at most by a factor
which can be uniformly bounded in energy:
$$
\rho_{\lambda,N}([E_0,E_1])\;\geq\;
C_0\;\lim_{L\to\infty}\;\int^{E_1}_{E_0}dE \int_{I_N}\PP(dv) \;
\frac{1}{\hat{R}_{\lambda,L}^E(v)^2}\; .
$$
Now we split into two contributions as in \eqref{eq-splitting}:
\begin{equation}
\label{eq-start0} \rho_{\lambda,N}([E_0,E_1])\;\geq\;
C_0\;\liminf_{L\to\infty}\;\int^{E_1}_{E_0}dE \int_{I_N}\PP(dv) \;
\frac{1}{\hat{R}^{E}_{\lambda,L,N}(\hat{\theta}^{E}_{\lambda,N}(v))^2}
\; \frac{1}{\hat{R}_{\lambda,N}^E(v)^2}\; .
\end{equation}
One may now use the positive constant (depending on $N$ and
$\lambda$, just as all the other constants  below as well)
$$
C_1\;=\;\max_{E_0\leq E\leq E_1,\;v\in
I_N}\;\hat{R}_{\lambda,N}^E(v)^2\;,
$$
in order to bound the second factor of the integrand in
\eqref{eq-start0}:
\begin{equation}
\label{eq-start} \rho_{\lambda,N}([E_0,E_1])\;\geq\;
\frac{C_0}{C_1}\;\liminf_{L\to\infty}\;\int^{E_1}_{E_0}dE
\int_{I_N}\PP(dv) \;
\frac{1}{\hat{R}^{E}_{\lambda,L,N}(\hat{\theta}^{E}_{\lambda,N}(v))^2}\;
.
\end{equation}
The strategy is to exhibit an adequate transformation of variables
in $I_N$ in order to be able to apply once again
Proposition~\ref{prop-alphaaverage}.

\vspace{.2cm}

First let us analyze which values the modified Pr\"ufer phase
$\hat{\theta}^{E}_{\lambda,N}(v)$ in \eqref{eq-start} may take. As
it is monotonous in each $v_n$ by
Proposition~\ref{prop-phasederivatives}{\rm (ii)} and because $m^E$
is a diffeomorphism, it follows that the smallest and largest values
are
$\hat{\theta}_0=\hat{\theta}_{\lambda,N}^E(-\frac{1}{2},\ldots,-\frac{1}{2})$
and
$\hat{\theta}_1=\hat{\theta}_{\lambda,N}^E(\frac{1}{2},\ldots,\frac{1}{2})$.
As the initial condition for the modified Pr\"ufer variables
is shifted by a term $\Oo(1)$ w.r.t. $N$,
one now has $\hat{\theta}_0=\hat{\theta}_{0,N}^{E+\frac{\lambda}{2}}(v)+\Oo(1)$ and
$\hat{\theta}_1=\hat{\theta}_{0,N}^{E-\frac{\lambda}{2}}(v)+\Oo(1)$, for arbitrary $v$.
Using \eqref{eq-modPr}, it therefore follows that
$$
\hat{\theta}_1-\hat{\theta}_0\;=\;
N\,\left[\arccos((E-\frac{\lambda}{2})/2)-\arccos((E+\frac{\lambda}{2})/2)\right]
\;+\;\Oo(1)\;\geq\; C_2N\lambda\;+\;\Oo(1)\;,
$$
so that one can choose $N$ of order $1/\lambda$ such that
$\hat{\theta}_1-\hat{\theta}_0>\pi$.

\vspace{.1cm}

For every $N\geq 2$,
$\|\nabla_v\hat{\theta}_{\lambda,N}^E(v)\|> 0$ by
Proposition~\ref{prop-phasederivatives}{\rm (ii)} because the
Pr\"ufer radius is bounded and no eigenfunction can vanish at two
consecutive sites. Therefore the map $v\in
I_N\mapsto\hat{\theta}_{\lambda,N}^E(v)$ has  no critical point and
the sets
\begin{equation}
\label{eq-submanifolds}
P_{\lambda,N}^E(\hat{\theta})\;=\; \left\{\;v\in
I_N\;\left|\;\hat{\theta}_{\lambda,N}^{E}(v)=\hat{\theta}\;\right. \right\}\;,
\qquad \hat{\theta}_0\leq \hat{\theta}\leq \hat{\theta}_1\;.
\end{equation}
are (real analytic) subvarieties of $I_N$ of co-dimension $1$, with
boundaries of co-dimension $2$. Then $[\hat{\theta}_0,\hat{\theta}_1]$ is
precisely the interval of $\hat{\theta}$'s for which
$P_{\lambda,N}^E(\hat{\theta})$ is not empty. Because the gradient (w.r.t. $v$)
of $\hat{\theta}_{\lambda,N}^E(v)$ does not vanish,
$P_{\lambda,N}^E(\hat{\theta}_0)$ and $P_{\lambda,N}^E(\hat{\theta}_1)$ consist
of only one point each and Morse theory implies that all other
manifolds $P^E_{\lambda,N}(\hat{\theta})$, $\hat{\theta}\in
(\hat{\theta}_0,\hat{\theta}_1)$, are diffeomorphic \cite[Theorem 6.2.2]{Hir}.
This implies also that the $N-1$-dimensional volume measured with
the $N-1$-dimensional Hausdorff measure $\Hh^{N-1}$ of the
hyper-surfaces $P_{\lambda,N}^E(\hat{\theta})$, $\hat{\theta}\in
(\hat{\theta}_0,\hat{\theta}_1)$, is positive. Therefore there exists a constant
$C_3>0$ such that $\Hh^{N-1}(P_{\lambda,N}^E(\hat{\theta}))\geq C_3$ for
all $\hat{\theta}$ in the smaller interval
$[\hat{\theta}_0',\hat{\theta}_1']\subset(\hat{\theta}_0,\hat{\theta}_1)$. This can be done
such that $\hat{\theta}_1'-\hat{\theta}_0'>\pi$.

\vspace{.1cm}

The change of variable will now be based on Federer's coarea formula
\cite{EG}. In the situation relevant for our purposes, it states
that for all Lipshitz continuous functions $g$
$$
\int_{I_N} \PP(dv) \;J_1(\hat{\theta}^{E}_{\lambda,N}(v))
\;g(\hat{\theta}^{E}_{\lambda,N}(v))\;=\; \int_{\hat{\theta}_0}^{\hat{\theta}_1}
d\hat{\theta}\;\mathcal{H}^{N-1}(P^E_{\lambda,N}(\hat{\theta}))\;g(\hat{\theta}) \;,
$$
where the 1-Jacobian is given by
$$
J_1(\hat{\theta}^{E}_{\lambda,N}(v))
\;=\;\left\|\nabla_v\hat{\theta}^{E}_{\lambda,N}(v)\right\|\;=\;
\left(\sum_{n=1}^N|\partial_{v_n}
\hat{\theta}^{E}_{\lambda,N}(v)|^2\right)^{\frac{1}{2}}\;.
$$
By compactness, one has $J_1(\hat{\theta}^{E}_{\lambda,N}(v))\leq
C_4$ for all $v\in I_N$ and $E\in [E_0,E_1]$. Now using the coarea
formula for $g(\hat{\theta})=\hat{R}^E_{\lambda,L,N}(\hat{\theta})^{-2}$ gives
\begin{eqnarray*}
\int_{I_N} \PP(dv)
\frac{1}{\hat{R}^{E}_{\lambda,L,N}(\hat{\theta}^{E}_{\lambda,N}(v))^2}
& \geq & \frac{1}{C_4}\; \int_{I_N}
\PP(dv)\;J_1(\hat{\theta}^{E}_{\lambda,N}(v))\;
\frac{1}{\hat{R}^{E}_{\lambda,L,N}(\hat{\theta}^{E}_{\lambda,N}(v))^2}
\\
& = & \frac{1}{C_4}\; \int_{\hat{\theta}_0}^{\hat{\theta}_1}
d\hat{\theta}\;\mathcal{H}^{N-1}(P_{\lambda,N}^E(\hat{\theta}))\;
\frac{1}{\hat{R}^{E}_{\lambda,L,N}(\hat{\theta})^2}
\\
& \geq & \frac{C_3}{C_4}\; \int_{\hat{\theta}_0'}^{\hat{\theta}_1'}
d\hat{\theta}\;\;\frac{1}{\hat{R}^{E}_{\lambda,L,N}(\hat{\theta})^2} \;.
\end{eqnarray*}
As $\hat{\theta}_1'-\hat{\theta}_0'>\pi$, the full projective space is covered. The M\"obius
transformation with $M^E$ does not change this property. Thus one
may pass back to non-modified
Pr\"ufer variables at the cost of another constant. Therefore the r.h.s.
can then be bounded below by a positive constant independent of $L$ due to
Proposition~\ref{prop-alphaaverage}. Replacing this bound in
\eqref{eq-start} completes the proof.
\hfill $\Box$

\vspace{.2cm}

We conclude this paragraph with some heuristics as to how the above proof can be modified in
order to yield an improvement of Wegner's lower bound \cite{Weg,HM}. More precisely, we shall argue that
one should be able to choose
$N=N(\lambda)=C_5\lambda^{-2}$ in Theorem~\ref{theo-Wegner} for
some adequate constant $C_5$ and that for this choice the lower bound on the averaged
spectral measure remains positive in the limit $\lambda\to 0$. The choice
$N=C_5\lambda^{-2}$ means that one is precisely at the scale of the localization
length \cite{PF,JSS}. Thus the norm of the transfer matrices still has not begun to grow
exponentially, and therefore $\hat{R}_{\lambda,N}^E(v)=\Oo(1)$ with high probability
in $v$ w.r.t. $\PP$. If one supposes that this holds uniformly in $v$ (which is wrong, of course,
and would have to be replaced by a probabilistic argument), one hence has $C_1=\Oo(1)$.
As $C_0$ is independent of $N$ and $\lambda$, this allows to start from \eqref{eq-start} with constants
of order of unity. Now due to the choice of $N(\lambda)$, one has
$\hat{\theta}_1-\hat{\theta}_0=\Oo(\lambda^{-\frac 1 2})$ by the same calculation as above,
which is much larger than the one turn
needed in order to conclude the argument. However, the $N-1$-dimensional volume of
$P_{\lambda,N}^E(\hat{\theta})$ is very small for most
$\hat{\theta}\in[\hat{\theta}_0,\hat{\theta}_1]$ and one has to select those
for which it is of order of unity. For this purpose, let us
recall from \cite{JSS} that the modified Pr\"ufer phases can be expanded as follows
\begin{equation}
\label{eq-Pexpan}
\hat{\theta}^{E}_{\lambda,N}(v)
\;=\;
\hat{\theta}^E_0\,+\,N\,k\,+\,
\lambda\sum_{n=1}^{N}v_n\left[1+\cos(2\hat{\theta}^{E}_{\lambda,n-1}(v))\right]
\;+\;\Oo(N\lambda^2)
\;.
\end{equation}
As the $v_n$ are centered and $N=C_5\lambda^{-2}$, the central limit theorem implies that
$\lambda\sum_{n=1}^{N}v_n$ converges in the limit $\lambda\to 0$
in distribution to a centered Gaussian. Hence this
term is of order $1$ with positive probability. On the other hand, the sum
$\lambda\sum_{n=1}^{N}v_n\cos(2\hat{\theta}^{E}_{\lambda,n-1}(v))$ is expected to
be of order $\lambda$ for almost all $v$. Neglecting this term as well as the error term
$\Oo(N\lambda^2)$ in \eqref{eq-Pexpan} then leads to
\begin{equation}
\label{eq-subm}
P_{\lambda,N}^E(\hat{\theta})\;\approx\; \left\{\;v\in
I_N\;\left|\;\hat{\theta}=\hat{\theta}^E_0\,+\,N\,k\,+\,
\lambda\sum_{n=1}^{N}v_n\;\right. \right\}\;.
\end{equation}
Now we choose $\hat{\theta}_0'=\hat{\theta}^E_0+Nk-C_6$ and $\hat{\theta}_1'=\hat{\theta}^E_0+Nk+C_6$
for some $C_6>0$ sufficiently large. By \eqref{eq-subm} and the central limit theorem, one then has
$\Hh^{N-1}(P_{\lambda,N}^E(\hat{\theta}))\geq C_7$ for
all $\hat{\theta}\in [\hat{\theta}_0',\hat{\theta}_1']$. Using that also the $1$-Jacobian is of
order $1$ (which can roughly be deduced from \eqref{eq-Pexpan}), this should allow to conclude the proof
just as above. The lower bound obtained is then independent of $\lambda$.

\section{Applications to spectral analysis}
\label{sec-applic}

The two applications of this section both concern the model \eqref{eq-mod2}, but the
first one is based on an application also of Theorem \ref{theo-secondexample}. Because
the appearing $J_N$ is arbitrary, the operators $H_{\lambda,N,v}$ can have all spectral
types. We state the following propositions for the singular part of the spectrum, but it
holds for many other case (see the Remark below).

\begin{proposi}
\label{prop-firstapp} Let us fix $\hat{\lambda}\in (0,4)$ as well as a
corresponding interval $\hat{I}$ and integer
$\hat{N}=N(\hat{\lambda})$ as given in {\rm Theorem \ref{theo-Wegner}}.
Furthermore, fix some
$\hat{v}=(\hat{v}_1,\ldots,\hat{v}_{\hat{N}})$ with $\hat{v}_n\geq 0$ and two adjacent
strictly positive entries. Then the following two statements are equivalent:

\vspace{.1cm}

\noindent {\rm (i)} $H_{\lambda,\hat{N},\hat{v}}$ has singular spectrum in
$\hat{I}$ for all $\lambda\in B$ where $B\subset \RR$ is of positive

Lebesgue measure, namely $|B|>0$.

\vspace{.1cm}

\noindent {\rm (ii)} $H_{\hat{\lambda},\hat{N},{v}}$ has singular spectrum in
$\hat{I}$ for all $v\in D\subset I_{\hat{N}}$ where $\PP(D)>0$.

\end{proposi}

\noindent {\bf Proof.} (i) $\Rightarrow$ (ii): Let $S\subset \RR$  be the set of
energies for which subordinate solutions of $H_{\lambda,\hat{N},\hat{v}}$ exist. This
set is a support of the singular part and is independent of $\lambda,\hat{N},\hat{v}$. If
$H_{\lambda,\hat{N},\hat{v}}$ has singular spectrum in $\hat{I}$
for $\lambda\in B$ where
$|B|>0$, then  one has
$\rho_{\lambda,\hat{N},\hat{v}}(S\cap \hat{I})>0$ for all $\lambda\in B$ (see {\it e.g.} by Corollary 2.8 of \cite{dRSS}). This implies
that $\int_\RR d\lambda \,\rho_{\lambda,\hat{N},\hat{v}}(S\cap \hat{I})>0$. Therefore
Theorem \ref{theo-secondexample} implies that $|S\cap \hat{I}|>0$. This in turn implies by
Theorem \ref{theo-Wegner} that
$\int_{I_N} \PP(dv) \,\rho_{\hat{\lambda},\hat{N},v}(S\cap \hat{I})>0$ so that
$\rho_{\hat{\lambda},\hat{N},v}(S\cap \hat{I})>0$ for all $v\in D$ with $\PP(D)>0$.
Hence again $H_{\hat{\lambda},\hat{N},v}$ has singular
spectrum in $\hat{I}$ for $v\in D$ with $\PP(D)>0$.

(ii) $\Rightarrow$ (i): If
$H_{\lambda,\hat{N},\hat{v}}$ has singular spectrum for a set of $v$'s of
positive measure, then
$\int_{I_N} \PP(dv) \,\rho_{\hat{\lambda},\hat{N},v}(S\cap \hat{I})>0$ so that
$|S\cap \hat{I}|>0$ by Theorem \ref{theo-Wegner}. For $\lambda_0$ and $\lambda_1$
adequately chosen, Theorem \ref{theo-secondexample} then implies that
$\int_{\lambda_0}^{\lambda_1} d\lambda \,\rho_{\lambda,\hat{N},\hat{v}}(S\cap \hat{I})>0$,
which shows (i).
\hfill $\Box$

\vspace{.2cm}

By the same proof evoking either Theorem \ref{theo-secondexample} or Theorem \ref{theo-Wegner} twice (and nowhere the other) one proves the following
results:

\begin{proposi}
\label{prop-secondapp2} Let us fix
an open interval $\hat{I}$ and integers
$\hat{N}_0,\hat{N}_1$ 
as well as positive $\hat{v}_0\in I_{\hat{N}_0},\hat{v}_1\in I_{\hat{N}_1}$.
Then the following two statements are equivalent:

\vspace{.1cm}

\noindent {\rm (i)} $H_{{\lambda},\hat{N}_0,\hat{v}_0}$ has singular
spectrum in
$\hat{I}$ for a set of $\lambda$'s of positive Lebesgue measure.

\vspace{.1cm}

\noindent {\rm (ii)} $H_{{\lambda},\hat{N}_1,\hat{v}_1}$ has singular
spectrum in
$\hat{I}$ for a set of $\lambda$'s of positive Lebesgue measure.

\end{proposi}

\begin{proposi}
\label{prop-secondapp} Let us fix
$\hat{\lambda}_0,\hat{\lambda}_1\in (0,4)$, and corresponding to both
as in {\rm Theorem \ref{theo-Wegner}}
an interval $\hat{I}$ and an integer
$\hat{N}=N(\hat{\lambda})$.
Then the following two statements are equivalent:

\vspace{.1cm}

\noindent {\rm (i)} $H_{\hat{\lambda}_0,\hat{N},{v}}$ has singular
spectrum in
$\hat{I}$ for all $v\in D_0\subset I_{\hat{N}}$ where $\PP(D_0)>0$.

\vspace{.1cm}

\noindent {\rm (ii)} $H_{\hat{\lambda}_1,\hat{N},{v}}$ has singular
spectrum in
$\hat{I}$ for all $v\in D_1\subset I_{\hat{N}}$ where $\PP(D_1)>0$.

\end{proposi}

\vspace{.2cm}

\noindent {\bf Remark:} The same results hold for the pure-point part of
the spectrum and its singular continuous part, as well as for
$\alpha$-continuity and $\alpha$-singularity of the spectral measures
(for the proof of the latter, combine the above arguments with
those of \cite{JL,KLS}.). With some care, it is possible to further
localize the set $B$ in Proposition \ref{prop-firstapp} in $(0,2)$.

\vspace{.2cm}

With a similar proof working with zero measure sets,
one can also obtain results analogous to the above propositions. We have, for example, the following:

\vspace{.2cm}

\begin{proposi}
\label{prop-thirdapp} Under the same hypothesis as in {\rm Proposition
\ref{prop-firstapp}},
the following two statements are equivalent:

\vspace{.1cm}

\noindent {\rm (i)} $H_{\lambda,\hat{N},\hat{v}}$ has pure-point spectrum in
$\hat{I}$ for Lebesgue almost all $\lambda\in \RR$.

\vspace{.1cm}

\noindent {\rm (ii)} $H_{\hat{\lambda},\hat{N},{v}}$ has pure-point spectrum in
$\hat{I}$ for $\PP$-almost all $v\in I_{\hat{N}}$.

\end{proposi}

It seems to be an open question whether this proposition is true when the word "pure" is omitted.

\vspace{.2cm}

\noindent {\bf Acknowledgment:}
R. del R. thanks B. Simon for pertinent hints to the literature.
Our work and respective visits were supported by PROYECTO PAPIIT IN-111906 UNAM and by the DFG. C. M. was funded by CONACYT.


\end{document}